\documentclass[conference,10pt]{IEEEtran}

\usepackage{algorithm}
\usepackage[noend]{algpseudocode}
\usepackage{bm}
\usepackage{stackrel}
\usepackage{subfigure}
\usepackage{bbm}
\usepackage{epsf}
\usepackage{amsmath,amssymb}
\usepackage{graphicx}
\usepackage{color}
\usepackage{cite}
\usepackage{multirow,tabularx}
\usepackage{ifthen}
\usepackage{epstopdf}
\input{epstopdf.sty}
\usepackage{framed}
\usepackage{times}

\input{epsf.sty}
\makeatletter
\def\BState{\State\hskip-\ALG@thistlm}
\makeatother

\newcommand{\hide}[1]{\ifthenelse{\boolean{false}}{#1}{}}

\newtheorem{theorem}{{\bf Theorem}}

\newtheorem{lemma}{{\bf Lemma}}

\newcommand{\qed}{\nobreak \ifvmode \relax \else
      \ifdim\lastskip<1.5em \hskip-\lastskip
      \hskip1.5em plus0em minus0.5em \fi \nobreak
      \vrule height0.75em width0.5em depth0.25em\fi}

\newcommand{\beq}{\begin{equation}}
\newcommand{\eeq}{\end{equation}}
\newcommand{\barr}{\begin{array}}
\newcommand{\earr}{\end{array}}

\newcommand{\benum}{\begin{enumerate}}
\newcommand{\eenum}{\end{enumerate}}

\newcommand{\bit}{\begin{itemize}}
\newcommand{\eit}{\end{itemize}}

\newcommand{\bc}{\begin{center}}
\newcommand{\ec}{\end{center}}

\newcommand{\bdes}{\begin{description}}
\newcommand{\edes}{\end{description}}

\newcommand{\bfig}{\begin{figure}}
\newcommand{\efig}{\end{figure}}

\newcommand{\bemq}{\begin{quote} \begin{em}}
\newcommand{\eemq}{\end{em} \end{quote}}

\newcommand{\bmp}{\begin{minipage}}
\newcommand{\emp}{\end{minipage}}

\newcommand{\apndx}[1]{Appendix~\ref{#1}}

\newcommand{\thmref}[1]{Theorem~\ref{#1}}

\newcommand{\supth}{^{{\mathrm{th}}}}

\newcommand{\EX}[1]{\mathbb{E}\left[{#1}\right]} % expectation operator

\newcommand{\bsp}{\begin{slide*}}
\newcommand{\esp}{\end{slide*}}
\newcommand{\bsl}{\begin{slide}}
\newcommand{\esl}{\end{slide}}

\newcommand{\blem}{\begin{lemma}}
\newcommand{\elem}{\end{lemma}}
\newcommand{\bthm}{\begin{theorem}}
\newcommand{\ethm}{\end{theorem}}

\newcommand{\pr}[1]{\mathbf{P}\left[ #1 \right]}

\IEEEoverridecommandlockouts

\begin{document}
\pdfoutput=1
\title{Age of Information for Discrete Time Queues}
% \date{\today}
\author{Vishrant Tripathi, Rajat Talak, and Eytan Modiano\\
Laboratory for Information \& Decision Systems, MIT\\
vishrant@mit.edu, talak@mit.edu, modiano@mit.edu}
%\thanks{This work has been submitted to the Age of Information Workshop (AoI'19), which will happen in conjunction with IEEE Infocom 2019.}}
% \thanks{The authors are with the Dept.\ of Electrical Communication Eng.\ in the Indian Institute of Science (IISc), Bangalore, India.}

\IEEEaftertitletext{\vspace{-0.6\baselineskip}}

\maketitle

\begin{abstract}
Age of information (AoI) is a time-evolving measure of information freshness, that tracks the time since the last received fresh update was generated. Analyzing peak and average AoI, two time average metrics of AoI, for various continuous time queueing systems has received considerable attention.
We analyze peak and average age for various discrete time queueing systems. We first consider first come first serve (FCFS) Ber/G/1 and Ber/G/1 queue with vacations, and derive explicit expressions for peak and average age.
We also obtain age expressions for the last come first serve (LCFS) queue and the $G/G/\infty$ queue. We build upon proof techniques from earlier results, and also present new techniques that might be of independent interest in analyzing age in discrete time queuing systems.
\end{abstract}

\section{Introduction}
A source generates updates, which traverse a network to reach the destination. The goal of the system designer is to ensure that the destination gets fresh information. Age of information (AoI), a destination centric metric of information freshness, was first introduced in \cite{2012Infocom_KaulYates}. It measures the time that elapsed since the last received fresh update was generated at the source. Over the past few years, a rapidly growing body of work has analyzed AoI for various queuing systems~\cite{2012Infocom_KaulYates, 2014ISIT_KamKomEp, 2014ISIT_CostaEp, 2016X_Najm, sun_lcfs_better, 2016_ISIT_YinSun_AoI_Thput_Delay_LCFS, Inoue17_FCFS_AoIDist, 2018_Ulukus_GG11, 2018ISIT_Yates_AoI_ParallelLCFS, 2011SeCON_Kaul, 2016_MILCOM_Ep_AoI_Buffer_Deadline_Replace, 2016_ISIT_Ep_AoI_Deadlines, 2018_ISIT_Inoue_AoI_Deadline} and wireless networks~\cite{2016allerton_IgorAge, 2016Ep_WiOpt, talak17_allerton, talak18_Mobihoc, talak18_WiOpt, 2012ISIT_YatesKaul}.

AoI was first studied for the first come first serve (FCFS) M/M/1, M/D/1, and D/M/1 queues in~\cite{2012Infocom_KaulYates}. AoI for M/M/2 and M/M/$\infty$ was studied in~\cite{2014ISIT_KamKomEp, 2014ISIT_CostaEp}, in order to demonstrate the advantage of having parallel servers. In~\cite{2018ISIT_Yates_AoI_ParallelLCFS}, age was analyzed for parallel last come first serve (LCFS) servers, with preemptive service. Age analysis for queues with packet deadlines, in which a packet deletes itself after its deadline expiration, is considered in~\cite{2016_MILCOM_Ep_AoI_Buffer_Deadline_Replace, 2016_ISIT_Ep_AoI_Deadlines, 2018_ISIT_Inoue_AoI_Deadline}. In~\cite{2016X_LongBo}, age has been analyzed under packet transmission errors. In~\cite{2016X_Najm}, AoI for the LCFS queue with Poisson arrivals and Gamma distributed service was considered.
In~\cite{sun_lcfs_better, 2016_ISIT_YinSun_AoI_Thput_Delay_LCFS}, the LCFS queue scheduling discipline, with preemptive service, is shown to be age optimal, when the service times are exponentially distributed.

More recently, a complete characterization of age distribution for FCFS and LCFS queues, with and without preemption, was done in~\cite{Inoue18_FCFS_LCFS_AoIDist}. In~\cite{talak18_determinacy}, it is proved that a heavy tailed service minimizes age for LCFS queue under preemptive service and the G/G/$\infty$ queue. Its extension~\cite{talak19_AoI_age_delay} proves an important age-delay tradeoff in single server systems.

AoI has thus far been analyzed for continuous time queuing models. Discrete time queuing systems often arise in practice, especially in wireless networks~\cite{talak18_Mobihoc}. In~\cite{talak18_Mobihoc}, we derived peak and average age expressions for the FCFS G/Ber/1 queue. The result lead to the derivation of separation principle in scheduling and rate control for age minimization in wireless networks. In this work, we analyze age metrics for various discrete time queuing models. We first consider the FCFS Ber/G/1 queue, with and without vacations. When taking vacations, we note that taking deterministic vacations, is the best resort towards minimizing age.

We then derive peak and average age expressions for the LCFS G/G/1 with preemptive service, and the infinite server G/G/$\infty$. We build upon our proof techniques from earlier results~\cite{Vishrant19_AoI_on_graph, talak18_determinacy, talak18_Mobihoc}, and also present new techniques that might be of independent interest in analyzing age in discrete time queuing systems.

\section{Age of Information}

We assume a slotted system with packets generated by a source according to a random process. We assume a service system consisting of one or more servers, depending on the setup, and packets taking random integer number of time-slots to get served. For analysis, we assume that packets arrive at the beginning of the time-slot and finish service at the end of the time-slot. We specify the inter-arrival distributions, service distributions and service disciplines in every section.

We track the age process $A(t)$ as the value of AoI at the beginning of every time-slot. Assume that the $i$th packet is generated at time $Y_i$. Then, $A(t)$ satisfies the following recursion
\begin{equation*}
A(t+1) =
 \begin{cases}
      A(t)+1, & \text{if no service at time }t \\
      \min \{ t-Y_i,A(t) \} +1, &  \text{if }i\text{ is served.}
 \end{cases}
\end{equation*}

See Figure~\ref{fig:lcfs} for an example age evolution plot.
Both peak and average age are defined as usual. The peak age $A^\text{p}$ is the time average of age values at time instants when there is useful packet delivery. The average age $A^\text{ave}$ is the time-average of the entire age process $A(t)$. Note that when a useful packet delivery occurs in time-slot $t$, then $A(t+1) \leq A(t)$. Thus,

\begin{equation}
A^{\text{p}} \triangleq \limsup_{T \rightarrow \infty} \frac{ \sum\limits_{t = 1}^{t = T} A(t)\mathbbm{1}_{ \{ A(t+1) \leq A(t)\}} }{ \sum\limits_{t = 1}^{t = T} \mathbbm{1}_{ \{ A(t+1) \leq A(t)\}} }, \text{ and}
\end{equation}

\begin{equation}
A^{\text{ave}} \triangleq \limsup_{T \rightarrow \infty} \frac{1}{T} \sum_{t = 1}^{T} A(t).
\end{equation}

We provide closed form expressions for $A^{\text{p}}$ and $A^{\text{ave}}$ for different queuing models.

\section{Ber/G/1 Queue}

Consider a discrete time Ber/G/1 queue, where an arrival occurs at time $t$ with probability $\lambda$, while the service times $S$ are generally distributed with mean $\EX{S} = 1/\mu$. We obtain expressions for peak and average age for the discrete time Ber/G/1 queue.
\begin{framed}
\begin{theorem}
\label{thm:bg1}
The peak and average age for the discrete time Ber/G/1 queue are given by
\begin{equation}
\label{eq:bg1_peak}
A^{\text{p}} = \frac{1}{\lambda} + \frac{1}{\mu} + \frac{\lambda \mathbb{E}[S^2]-\rho}{2(1-\rho)},
\end{equation}
and
\begin{equation}
\label{eq:bg1_ave}
A^{\text{ave}} = 1 + \frac{1}{\mu} +  \frac{(1-\lambda)(1-\rho)}{\lambda L_S(1-\lambda)} + \frac{\lambda \mathbb{E} [S^2] - \rho}{2(1-\rho)},
\end{equation}
where $L_S(x) \triangleq \mathbb{E} [x^S]$ is the probability generating function of $S$ and $\rho  = \frac{\lambda}{\mu}$.
\end{theorem}
\end{framed}
\begin{IEEEproof}
The peak age for an FCFS queue is given by \cite{huang_peak}
\begin{equation}\label{eq:bg1_1}
A^{\text{p}} = \EX{T + X},
\end{equation}
where $T$ denotes the time an update sends in the queue and $X$ is the inter-arrival time between two updates. From~\cite[Chapter~4.6.1]{bose2013introduction}, for a Ber/G/1 queue we have
\begin{equation}
\label{eq:bg1_2}
\mathbb{E}[T] = \frac{\lambda \mathbb{E}[S^2]-\rho}{2(1-\rho)} + \frac{1}{\mu},
\end{equation}
where $S$ denotes the service time. Substituting this and $\EX{X} = \frac{1}{\lambda}$ in~\eqref{eq:bg1_1}, we obtain the expression for peak age. For the derivation of average age see Appendix~\ref{pf:bg1_avg}.
\end{IEEEproof}
We observe that the peak age expression for a Ber/G/1 queue is near identical to that of the M/G/1 queue derived in \cite{talak18_determinacy} with an additional term $\frac{-\rho}{2(1-\rho)}$ added due to the discretization. We use the probability generating function for analyzing average age due to the discrete nature of the service distribution.

\section{Ber/G/1 Queue with Vacations}
Consider a discrete time Ber/G/1 queue with vacations, where an arrival occurs at time $t$ with probability $\lambda$, while the service times $S$ are generally distributed with mean $\EX{S} = 1/\mu$. When the queue is empty, the server takes i.i.d. vacations $V$ that are generally distributed with mean $\EX{V}$, until a new arrival enters the queue. Ber/G/1 queues with vacations were used to find age optimal random walks for information dissemination on graphs in \cite{Vishrant19_AoI_on_graph}. M/M/1 queues with vacations were also used to study the age of updates in a simple relay network in \cite{2018_Maatouk_AoI_Relay}.
We obtain an expression for peak age and bounds for average age in the FCFS discrete time Ber/G/1 queue with vacations.
\begin{framed}
\begin{theorem}
\label{thm:bgv}
The peak age for the discrete time Ber/G/1 queue with vacations is given by
\begin{equation}
\label{eq:bgv_peak}
A^{\text{p}} = \frac{1}{\lambda} + \frac{1}{\mu} + \frac{\lambda \mathbb{E}[S^2]-\rho}{2(1-\rho)} + \frac{\EX{V^2}}{2\EX{V}}-\frac{1}{2},
\end{equation}
and the average age is upper bounded by the peak age
\begin{equation}
\label{eq:buff_ave}
A^{\text{ave}} \leq A^{\text{p}}.
\end{equation}
%where $L_S(x) \triangleq \mathbb{E} [x^S]$ is the probability generating function of $S$ and $\rho  = \frac{\lambda}{\mu}$.
\end{theorem}
\end{framed}
\begin{IEEEproof}
As usual, the peak age for an FCFS queue is given by \cite{huang_peak}
\begin{equation}\label{eq:bgv_1}
A^{\text{p}} = \EX{T + X}.
\end{equation}
%where $T$ denotes the time an update sends in the queue and $X$ is the inter-arrival time between two updates.
Given that vacation times are distributed i.i.d according to random variable $V$, using a residual time argument one can show that \cite{tian2006vacation}
\begin{equation}
\label{eq:bgv_2}
\mathbb{E}[T] = \frac{\lambda \mathbb{E}[S^2]-\rho}{2(1-\rho)} + \frac{1}{\mu} + \frac{\EX{V^2}}{2\EX{V}}-\frac{1}{2},
\end{equation}
where $S$ denotes the service time. Substituting this and $\EX{X} = \frac{1}{\lambda}$ in~\eqref{eq:bgv_1}, we obtain the expression for peak age.

\begin{figure}
  \centering
  \includegraphics[width=\linewidth]{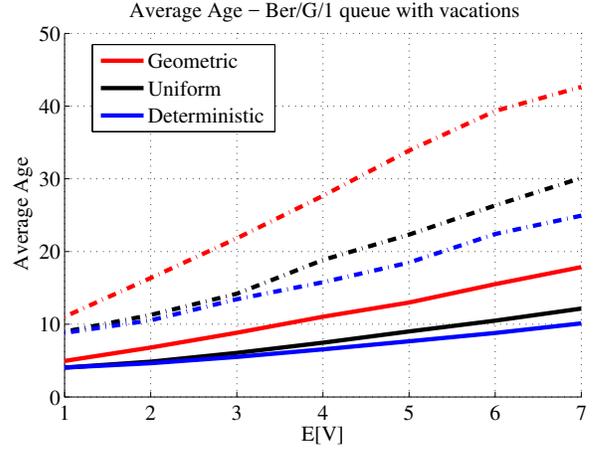}
  \caption{Averag age for geometrically distributed server times with probability $\mu = 0.75$. We compare vacations with geometric, uniform bounded, and deterministic distributions having the same mean as it varies from 1 to 7. Solid lines represent arrival probability $\lambda = 0.3$ and dashed lines represent $\lambda = 0.6$.}\label{fig:bgv_avg}
  \vspace{-0.0em}
\end{figure}

For a stable FCFS discrete time queue with Bernoulli arrivals, the average age is given by \cite{2012Infocom_KaulYates}
\begin{equation}
A^{\text{ave}} = \frac{1}{\lambda} + \lambda \EX{X_n T_n},
\end{equation}
where $X_1, X_2, ...$ are i.i.d. packet inter-arrival times and $T_1, T_2,...$ are  corresponding times spent in the system by each packet. Observe that $X_n$ and $T_n$ are negatively correlated - a smaller inter-arrival time means more congestion and more time spent in the system. Thus, \begin{equation}
A^{\text{ave}} \leq \frac{1}{\lambda} + \lambda \EX{X_n}\EX{T_n} = \EX{X_n} + \EX{T_n} = A^{\text{p}},
\end{equation}
i.e. the average age is upper bounded by the peak age of the system. For derivation of tighter upper and lower bounds on average age for a Ber/G/1 queue with vacations see Appendix~\ref{pf:bgv_avg}.
\end{IEEEproof}

We observe that the peak age for a Ber/G/1 queue with vacations splits into two terms - the peak age for a Ber/G/1 queue without vacations, as derived in the previous section, and a term that depends only on the vacations. From Figure \ref{fig:bgv_avg}, we also observe numerically that the lighter the tail of the vacation distribution, better the age. We see that deterministic vacations minimize average age, given a fixed value of $\EX{V}$.

%\section{PH Arrival and Service Times}

\section{LCFS Queues}
\label{sec:lcfs}
Consider a discrete time LCFS G/G/1 queue with preemptive service, in which a newly arrived packet gets priority for service immediately. We assume that packets arrive at the beginning of a time-slot and leave at the end of a time-slot. Update packets are generated according to a renewal process, with inter-generation times distributed according to $p_{X}$. The service times are distributed according to $p_{S}$, i.i.d. across packets. We derive explicit expressions for peak and average age for general inter-generation and service time distributions. %Note that both inter-arrival times and service times are discrete valued random variables.

\begin{figure}
  \centering
  \includegraphics[width=0.95\linewidth]{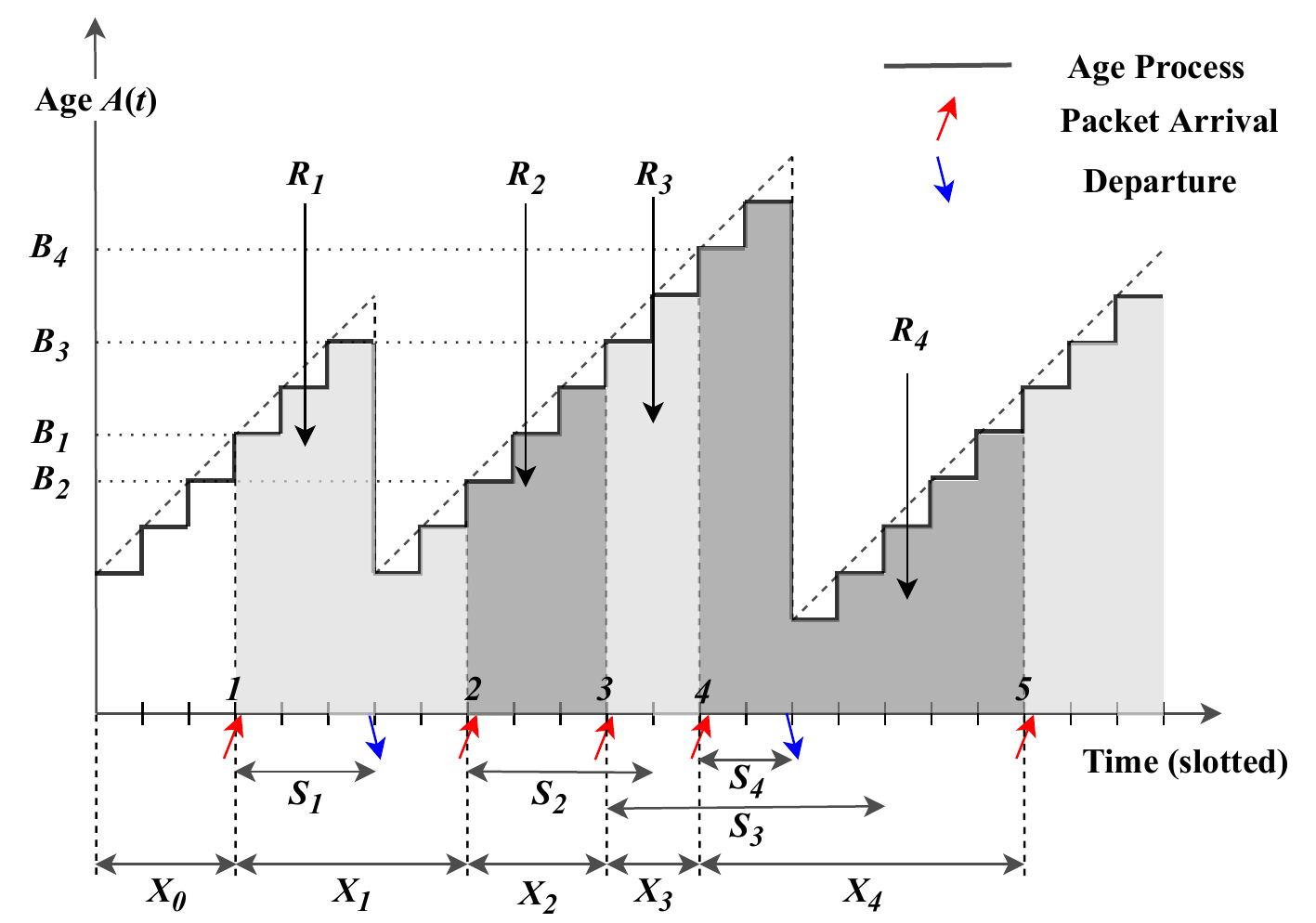}
  \caption{Age $A(t)$ evolution in time $t$ for the LCFS queue with preemption.}\label{fig:lcfs}
  \vspace{-0.0em}
\end{figure}
Let $X_i$ denote the inter-generation time between the $i$th and $(i+1)$th update packet. Due to preemption, not all packets get serviced on time to contribute to age reduction. We illustrate this in Figure~\ref{fig:lcfs}. Observe that packets $2$ and $3$ arrive before packet $4$. However, packet $2$ is preempted by packet $3$, which is subsequently preempted by packet $4$. Thus, packet $4$ is serviced before $2$ and $3$. Service of packet $2$ and $3$ (not shown in figure) does not contribute to age curve $A(t)$ because they contain stale information.
\begin{framed}
\begin{theorem}
\label{thm:LCFS_gg1}
For the discrete time LCFS G/G/1 queue, the peak and average age are given by
\begin{equation}\nonumber
A^{\text{p}}_{\text{G/G/1}} = \frac{\EX{X}}{\pr{S \leq X}} + \frac{\EX{S\mathbb{I}_{S \leq X}}}{\pr{S \leq X}}-1,
\end{equation}
and
\begin{equation}\nonumber
A^{\text{ave}}_{\text{G/G/1}} = \frac{1}{2}\frac{\EX{X^2}}{\EX{X}} + \frac{\EX{\min\left(X, S\right) }}{\pr{S \leq X}} - \frac{1}{2},
\end{equation}
where $X$ and $S$ denotes the independent inter-generation and service time random variables, respectively.
\end{theorem}
\end{framed}

\begin{IEEEproof}
\textbf{1. Peak Age:}
Let $A(t)$ denote the age at time $t$. Let $B_i$ denote the age at the generation of the $i$th update packet, i.e. $Z_i = \sum_{k=0}^{i-1}X_k$:
\begin{equation}
B_{i} = A( Z_i ).
\end{equation}
Then, we have the following recursion for $B_i$:
\begin{equation}
B_{i+1} = \left\{ \begin{array}{cc}
                    X_i & \text{if}~S_i \leq X_i \\
                    B_i + X_i & \text{if}~S_i > X_i
                  \end{array}\right.,
\end{equation}
for all $i \geq 0$. This can be written as
\begin{equation}
B_{i+1} = X_i + B_{i}\left( 1 - \mathbb{I}_{S_i \leq X_i}\right).
\end{equation}
Note that $B_i$ is independent of $S_i$ and $X_i$. Further, $\{ B_i \}_{i \geq 1}$ is a Markov process, and can be shown to be positive recurrent using the drift criteria~\cite{meyn_markov_chains_stability}. Taking expected value, and noting that at stationarity $\EX{B_i} = \EX{B_{i+1}}$, we get
\begin{equation}\label{eq:B_exp}
\EX{B} = \frac{\EX{X}}{\pr{S \leq X}}.
\end{equation}

We now compute the peak age. Let $P_i$ denote the peak value at the $i$th virtual service defined to be:
\begin{equation}
P_i = A(Z_i + S_i-1) \mathbb{I}_{S_{i} \leq X_i},
\end{equation}
where the event $\{ S_i \leq X_i \}$ denotes that the $i$th update packet was services, and not preempted. Note that $P_i = 0$ otherwise. When $\{ S_i \leq X_i \}$, we have $A(Z_i + S_i-1) = A(Z_i) + S_i-1 = B_i + S_i-1$. Therefore,
\begin{equation}
P_i = \left(B_i + S_i-1\right)\mathbb{I}_{S_{i} \leq X_i}.
\end{equation}
Using ergodicity of $\{ B_i \}_{i \geq 1}$ we obtain
\begin{equation}
\label{eq:oo1}
\lim_{M \rightarrow \infty} \frac{1}{M}\sum_{i=1}^{M} P_i = \EX{B}\EX{\mathbb{I}_{S \leq X}} + \EX{(S-1) \mathbb{I}_{S \leq X}},
\end{equation}
since $B_i$ is independent of $X_i$ and $S_i$.
The peak age can be written as:
\begin{equation}
A^{\text{p}}_{\text{G/G/1}} = \lim_{M \rightarrow \infty} \EX{ \frac{\sum_{i=1}^{M} P_i}{\sum_{i=1}^{M}\mathbb{I}_{S_i \leq X_i}}}.
\end{equation}
Using~\eqref{eq:oo1}, and the strong law of large numbers in the denominator, we get:
\begin{equation}
A^{\text{p}}_{\text{G/G/1}} = \frac{\EX{B}\pr{S \leq X} + \EX{(S-1) \mathbb{I}_{S\leq X}}}{\pr{S \leq X}}.
\end{equation}
Substituting for $\EX{B}$ (from~\eqref{eq:B_exp}) we obtain:
\begin{equation}
A^{\text{p}}_{\text{G/G/1}} = \frac{\EX{X}}{\pr{S \leq X}} + \frac{\EX{S \mathbb{I}_{S \leq X}}}{\pr{S \leq X}} - 1.
\end{equation}

\textbf{2. Average Age:} We take a different approach to analyzing the average age. Let $R_i$ denote the area under the age curve $A(t)$ between the generation of packet $i$ and packet $i+1$:
\begin{equation}
R_i \triangleq \sum_{t = Z_i}^{Z_i + X_i-1} A(t) ,
\end{equation}
where $Z_i = \sum_{k=0}^{i-1}X_k$ is the time of generation of the $i$th update packet. This $R_i$ can be computed explicitly to be
\begin{equation}
R_i = \left\{ \begin{array}{cc}
                B_i X_i + \frac{1}{2}(X_{i}^{2}-X_i) & \text{if}~X_i < S_i \\
                B_i S_i + \frac{1}{2}(X_{i}^{2}-X_i) & \text{if}~X_i \geq S_i
              \end{array}\right.,
\end{equation}
which can be written compactly as
\begin{equation}\label{eq:Ri}
R_i = \frac{1}{2}(X^{2}_{i}-X_i) + B_{i}\min\left( X_i, S_i\right).
\end{equation}
Since, $B_i$ is independent of $X_i$ and $S_i$, taking expected value at stationarity we obtain
\begin{equation}\label{eq:R_exp}
\EX{R} = \frac{1}{2}\EX{X^2-X} + \EX{B}\EX{\min\left( X, S\right)}.
\end{equation}

Using renewal theory, the average age can be obtained to be
\begin{align}
A^{\text{ave}}_{\text{G/G/1}} &= \frac{\EX{R}}{\EX{X}} = \frac{1}{2}\frac{\EX{X^2}}{\EX{X}} + \frac{\EX{B}}{\EX{X}}\EX{\min\left( X, S\right)} - \frac{1}{2}.
\end{align}
Substituting~\eqref{eq:B_exp} we get the result.
\end{IEEEproof}

We again observe the similarity of peak and average ages in the continuous and discrete time cases. Compared to the expressions in \cite{talak18_determinacy}, the discrete time peak age has extra discretization term of $-1$, while the average age has a discretization term of $-\frac{1}{2}$. Also, note that the strict inequality in the age expression in \cite{talak18_determinacy} changes to a non-strict inequality for the discrete time queue.

\section{Infinite Servers}
\label{sec:inf_serv}
Next, consider the G/G/$\infty$ queue, where every newly generated packet is assigned a new server. Let $p_X$ and $p_S$ denote the pmfs of inter-generation and service times, respectively. We focus only on the average age metric, and leave the optimization of peak age for future work.
\begin{framed}
\begin{theorem}
\label{lem:gginf}
For a discrete time G/G/$\infty$ queue, the average is given by
%\begin{equation}
%\nonumber
%A^{\text{p}}_{\text{G/G/}\infty} = \EX{X} + \EX{\min_{l \geq 0}\left\{ \sum_{k=1}^{l}X_{k} + S_{l+1}\right\} },
%\end{equation}
%and
\begin{equation}
A^{\text{ave}}_{\text{G/G/}\infty} = \frac{1}{2}\frac{\EX{X^2}}{\EX{X}} + \EX{\min_{l \geq 0}\left\{ \sum_{k=1}^{l}X_{k} + S_{l+1}\right\} } - \frac{1}{2}, \nonumber
\end{equation}
where $X$ and $\{ X_{k} \}_{k \geq 1}$ are i.i.d. distributed according to $p_X$, while $\{ S_{k} \}_{k \geq 1}$ are i.i.d. distributed according to $p_{S}$.
\end{theorem}
\end{framed}
\begin{IEEEproof}
For the G/G/$\infty$ queue, each arriving packet is serviced by a different server. As a result, the packets may get serviced in an out of order fashion. Figure~\ref{fig:gginf}, which plots age evolution for the G/G/$\infty$ queue, illustrates this. In Figure~\ref{fig:gginf}, observe that packet $3$ completes service before packet $2$. As a result, the age doesn't drop at the service of packet $3$, as it now contains stale information. To analyze average age, it is important to characterize these events of out of order service.

Let $X_i$ denote the inter-generation time between the $i$th and $(i+1)$th packet, and $S_i$ denote the service time for the $i$th packet. In
Figure~\ref{fig:gginf}, $X_2 + S_3 < S_2$, and therefore, packet $3$ completes service before packet $2$. To completely characterize this,
define $Z_i \triangleq \sum_{k=0}^{i-1} X_k$ to be the time of generation of the $i$th packet.
Note that the $i$th packet gets serviced at time $Z_i + S_i-1$, and the age drop due to packet $i$ getting served, if any, happens at time $Z_i + S_i$. The $(i+1)$th packet causes an age drop $Z_i + X_i + S_{i+1}$, and similarly, the $(i+l)$th packet causes an age drop, if any, at time $Z_i + \sum_{k=1}^{l}X_{i+k-1} + S_{i+l}$, for all $l \geq 1$.
Let $D_i$ denote the time from the $i$th packet generation to the time there is a possible age drop due to the $i$th packet, or a packet that arrived after the $i$th packet, whichever comes first. Thus,
\begin{align}
D_i &= \min\{S_i, X_{i} + S_{i+1}, X_{i} + X_{i+1} + S_{i+2}, \ldots \} \nonumber \\
    &= \min_{l \geq 0}\left\{ \sum_{k=1}^{l}X_{i+k-1} + S_{i+l}\right\}.
\end{align}
In Figure~\ref{fig:gginf}, note that $D_1 = S_1$, $D_2 = X_2 + S_3$, $D_3 = S_3$, and $D_4 = S_4$.
\begin{figure}
  \centering
  \includegraphics[width=0.95\linewidth]{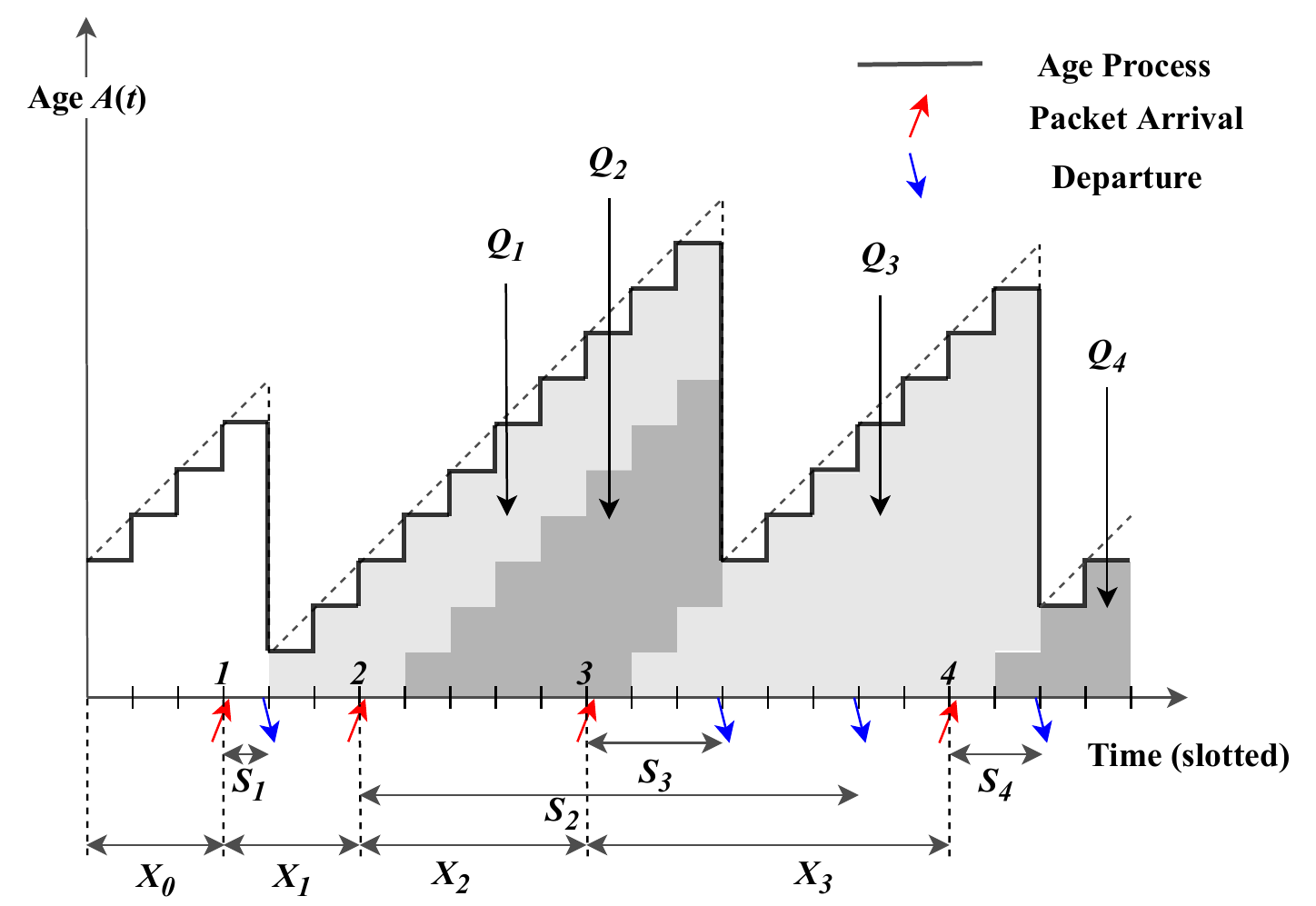}
  \caption{Age $A(t)$ evolution over time $t$ for G/G/$\infty$ queue.}
  \label{fig:gginf}
  \vspace{-0.0em}
\end{figure}
%We note that the $i$th peak is nothing but $X_i + D_{i+1}$. Therefore, the peak age is given by
%\begin{align}\nonumber
%A^{\text{p}}_{\text{G/G/}\infty} &= \EX{X_i} + \EX{D_{i+1}}, \\
%&= \EX{X_i} + \EX{\min_{l \geq 0}\left\{ \sum_{k=1}^{l}X_{i+k-1} + S_{i+l}\right\}}, \nonumber \\
%&= \EX{X} + \EX{\min_{l \geq 0}\left\{ \sum_{k=1}^{l}X_{k-1} + S_{l}\right\}}, \nonumber
%\end{align}
%since the $X_i$s and $S_i$s are independent and identically distributed.

The area under the age curve $A(t)$ is nothing but the sum of the areas of the regions $Q_i$ (see Figure~\ref{fig:gginf}). Applying the renewal reward theorem~\cite{wolff}, by letting the reward for the $i$th renewal, between $Z_i+1$ and $Z_i + X_i$, be the area $Q_i$, we get the average age to be:
\begin{equation}\label{eq:z0}
A^{\text{ave}}_{\text{G/G/}\infty} = \frac{\EX{Q_i}}{\EX{X_i}}.
\end{equation}
It is easy to see that
\begin{equation}
Q_i = \frac{(X_i + D_{i+1})(X_i + D_{i+1}-1)}{2} - \frac{D_{i+1}(D_{i+1}-1)}{2}. \label{eq:z1}
\end{equation}
% as the trapezoid $Q_i$ extends from the time of the $i$th packet generation to the time at which the $(i+1)$th, or a packet that arrives after the $(i+1)$th packet, is served; which is nothing but $X_{i} + D_{i+1}$.
%For illustration, note that $Q_1 = \frac{1}{2}(X_1 + X_2 + S_3)(X_1 + X_2 + S_3 -1) - \frac{1}{2}(X_2 + S_3)(X_2 + S_3-1)$, which is same as~\eqref{eq:z1}, for $i = 1$, since $D_2 = X_2 + S_3$.
Substituting~\eqref{eq:z1} in~\eqref{eq:z0}, we obtain
\color{black}
\begin{equation}
A^{\text{ave}}_{\text{G/G/}\infty} = \frac{1}{2}\frac{\EX{X^2}}{\EX{X}} + \frac{\EX{X_i D_{i+1}}}{\EX{X_i}}-\frac{1}{2}.
\end{equation}
We obtain the result by noting that $X_i$ and $D_{i+1}$ are independent.
%See Appendix~\ref{pf:lem:gginf}.
\end{IEEEproof}

We again observe a discretization factor of $-\frac{1}{2}$ as compared to the continuous time expression, derived in \cite{talak18_determinacy}.

\section{Conclusion}
\label{sec:conclusion}

Age of Information (AoI) is a time evolving measure of information freshness, that measures the time since the last received fresh update was generated at the source. AoI has mostly been analyzed only for continuous time queueing models. We analyse peak and average age for several discrete time queues. We first analyze peak and average age for the FCFS Ber/G/1 queue, with and without vacations, and see that taking deterministic vacations improves age, for a given mean vacation duration. We then derive peak and average age expressions for the LCFS G/G/1 queue, and the infinite server G/G/$\infty$ queue.
\bibliographystyle{ieeetr}
%\bibliography{../../../../PaperTrack/books-bib,../../../../PaperTrack/cvxalgo-bib,../../../../PaperTrack/aoi-bib,../../../../PaperTrack/uavnet-bib,../../../../PaperTrack/cps-bib,../../../../PaperTrack/neelesh-bib,../../../../PaperTrack/opt-scheduling-bib}

\appendix

\subsection{Proof for Average Age in \thmref{thm:bg1}}
\label{pf:bg1_avg}
Consider a Ber/G/1 queue with i.i.d. packet inter-arrival times $X_1, X_2, ...$ Let $T_n$ be the total time spent in the system by the $n\supth$ packet. Then, the average age is given by \cite{2012Infocom_KaulYates}
% \begin{equation}
% A^\text{ave} = \frac{1}{T} \sum\limits_{t = 1}^{t = T} A(t) = \frac{\sum\limits_{n=1}^{n=N} ((X_n+T_n)(X_n+T_n+1) - T_n(T_n+1))}{2\sum\limits_{n=1}^{n=N} X_n}.
% \end{equation}

% Dividing the numerator and denominator by $N$ and taking the limit as $N \rightarrow \infty$, we get -

\begin{equation}
A^\text{ave} = \frac{1}{\gamma} + \gamma \mathbb{E}[X_n T_n],
\end{equation}
where $\frac{1}{\gamma} = \mathbb{E}[X_n]$ and a packet arrives in every time-slot with probability $\gamma$. To evaluate the term $\mathbb{E}[X_n T_n]$, we use the following recursion -
\begin{equation}
T_n = \max \{T_{n-1}-X_n,0 \} + S_n,
\end{equation}
where $S_n$ is the service time of the $n\supth$ packet. Note that $T_{n-1}$ and $S_n$ are independent of $X_n$. Let $\EX{S_n} = \frac{1}{\mu}$ and $\rho \triangleq \frac{\gamma}{\mu}$. Evaluating $\EX{X_n T_n}$, we have
\begin{equation}
\label{eq:pf:bg1_1}
\begin{split}
\mathbb{E} [X_n T_n] & = \mathbb{E}[X_n \max \{T_{n-1}-X_n,0 \} ] + \mathbb{E}[S_n X_n], \\
%& = \mathbb{E}\bigg[ \mathbb{E}[X_n \max \{T_{n-1}-X_n,0 \} | T_{n-1}]\bigg] + \frac{\mathbb{E}[S]}{\gamma}, \\
& = \sum_{t=1}^{\infty} \mathbb{E}[X_n \max \{t-X_n,0 \}] \mathbb{P} (T = t) + \frac{\mathbb{E}[S]}{\gamma},
\end{split}
\end{equation}
where $\mathbb{P} (T = t)$ is the probability mass function of the total time spent by a packet in the system. We need to evaluate the term $\mathbb{E}[X_n \max \{t-X_n,0 \}].$
\begin{equation}
\label{eq:pf:bg1_2}
\begin{split}
\mathbb{E}[X_n \max \{t-X_n,0 \}]  = \sum_{x=1}^{t} x(t-x) \mathbb{P} (X_n = x), \\
 = \sum_{x=1}^{t} x(t-x) \gamma (1-\gamma)^{x-1},\\
= \frac{2(1-\gamma)^t - 2}{\gamma^2} + \frac{t(1-\gamma)^t - (1-\gamma)^t + t+1}{\gamma}.
\end{split}
\end{equation}

Using \eqref{eq:pf:bg1_1} and \eqref{eq:pf:bg1_2}, we now compute $\mathbb{E} [X_n T_n]$ as

\begin{equation}
\begin{split}
\mathbb{E} [X_n T_n] = \frac{2\mathbb{E}[(1-\gamma)^T] - 2}{\gamma^2} + \frac{\mathbb{E}[T(1-\gamma)^T]}{\gamma} + \\
\frac{\mathbb{E}[T]+1- \mathbb{E}[(1-\gamma)^T]}{\gamma} + \frac{\EX{S}}{\gamma}.
\end{split}
\end{equation}

We define $L_T(x) \triangleq \mathbb{E} [x^T]$. Then, $\mathbb{E}[(1-\gamma)^T] = L_T(1-\gamma),$ and $\mathbb{E}[T(1-\gamma)^T] = \frac{d}{dz} L_T(z) \big|_{z = 1-\gamma} (1-\gamma).$ Also, from \cite{bose2013introduction}, we know that for a Ber/G/1 queue, the probability generating function of $T$ is given by the following equation
\begin{equation}
L_T(z) = \frac{(1-\rho)(1-z)L_S(z)}{(1-z) - \gamma (1-L_S(z))}.
\end{equation}
Substituting $z = (1-\gamma)$ in the above expression we get %and differentiating we get $L'_T(1-\gamma)$. Simplifying, we obtain
\begin{equation}
\begin{split}
L_T(1-\gamma) & = 1-\rho, \text{ and }\\
\frac{d}{dz} L_T(z) \bigg|_{z = 1-\gamma} & = \frac{(1-\rho)}{\gamma}\bigg( \frac{1}{L_s(1-\gamma)}-1 \bigg).
\end{split}
\end{equation}
Putting all of these together along with the expression for $\EX{T}$, we get
\begin{equation}
A^\text{ave} = 1 + \mathbb{E}[S] +  \frac{(1-\gamma)(1-\rho)}{\gamma L_S(1-\gamma)} + \frac{\gamma \mathbb{E} [S^2] - \rho}{2(1-\gamma \mathbb{E}[S])}.
\end{equation}

\subsection{Bounds for Average Age in \thmref{thm:bgv}}
\label{pf:bgv_avg}
Consider a Ber/G/1 queue with i.i.d. packet inter-arrival times $X_1, X_2, ...$ and i.i.d. vacations $V_1, V_2, ...$ whenever the queue is empty. The proof is similar to the one in \apndx{pf:bg1_avg}. We modify the system time recursion as follows
%Let $T_n$ be the total time spent in the system by the $n\supth$ packet. Then, the average age is given by \cite{kaul2012real}
% \begin{equation}
% A^\text{ave} = \frac{1}{\gamma} + \gamma \mathbb{E}[X_n T_n],
% \end{equation}
% where $\frac{1}{\gamma} = \mathbb{E}[X_n]$ and a packet arrives in every time-slot with probability $\gamma$. To evaluate the term $\mathbb{E}[X_n T_n]$, we use the following recursion -
\begin{equation}
T_n = \max \{T_{n-1}-X_n,f_v(X_n-T_{n-1}) \} + S_n,
\end{equation}
where $S_n$ is the service time of the $n\supth$ packet and $f_v(X_n-T_{n-1})$ is the total random time the server spends in vacations if $X_n > T_{n-1}$ and zero otherwise. Note that $T_{n-1}$ and $S_n$ are independent of $X_n$. Let $\EX{S_n} = \frac{1}{\mu}$, $\EX{V_i} = v$, and $\rho \triangleq \frac{\gamma}{\mu}$. Evaluating $\EX{X_n T_n}$
\begin{equation*}
\begin{split}
& = \mathbb{E}[X_n \max \{T_{n-1}-X_n,f_v(X_n-T_{n-1}) \} ] + \mathbb{E}[S_n X_n], \\
%& = \mathbb{E}\bigg[ \mathbb{E}[X_n \max \{T_{n-1}-X_n,f_v(X_n-T_{n-1}) \} | T_{n-1}]\bigg] + \frac{\mathbb{E}[S]}{\gamma}, \\
& = \sum_{t=1}^{\infty} \mathbb{E}[X_n \max \{t-X_n,f_v(X_n-t) \}] \mathbb{P} (T = t) + \frac{\mathbb{E}[S]}{\gamma},
\end{split}
\end{equation*}
where $\mathbb{P} (T = t)$ is the probability mass function of the total time spent by a packet in the system. Evaluating the term $\mathbb{E}[X_n \max \{t-X_n,f_v(X_n-t) \}]$
\begin{equation*}
\begin{split}
& = \sum_{x=1}^{t} x(t-x) \mathbb{P} (X_n = x) + \sum_{x=t+1}^{\infty} x\bigg\lceil\frac{x-t}{v}\bigg\rceil v \mathbb{P} (X_n = x), \\
& \geq \sum_{x=1}^{t} x(t-x) \gamma (1-\gamma)^{x-1} + \sum_{x=t+1}^{\infty} x(x-t) \gamma (1-\gamma)^{x-1},\\
& = \frac{2(1-\gamma)^t - 2}{\gamma^2} + \frac{t(1-\gamma)^t - (1-\gamma)^t + t+1}{\gamma} \\
& + \frac{(1-\gamma)^t(\gamma t - \gamma + 2)}{\gamma^2}.
\end{split}
\end{equation*}
Note that $\EX{f_v(X_n-T_{n-1})|X_n = x, T_{n-1} = t} = \big\lceil \frac{x-t}{v} \big\rceil v$. We have used $\lceil a \rceil \geq a$ in the above analysis. Using this, we now lower bound $\mathbb{E} [X_n T_n]$ as

\begin{equation}
\label{pf:bgv:eq1}
\begin{split}
\mathbb{E} [X_n T_n] \geq \frac{4\mathbb{E}[(1-\gamma)^T] - 2}{\gamma^2} + \frac{2\mathbb{E}[T(1-\gamma)^T]}{\gamma} + \\
\frac{\mathbb{E}[T]+1- 2\mathbb{E}[(1-\gamma)^T]}{\gamma} + \frac{\EX{S}}{\gamma}.
\end{split}
\end{equation}

We define $L_T(x) \triangleq \mathbb{E} [x^T]$. Then, $\mathbb{E}[(1-\gamma)^T] = L_T(1-\gamma),$ and $\mathbb{E}[T(1-\gamma)^T] = \frac{d}{dz} L_T(z) \big|_{z = 1-\gamma} (1-\gamma).$ Also, from \cite{tian2006vacation}, we know that for a Ber/G/1 queue with vacations, the probability generating function of $T$ is given by the following equation
\begin{equation}
L_T(z) = \bigg( \frac{(1-\rho)(1-z)L_S(z)}{(1-z) - \gamma (1-L_S(z))}\bigg) \frac{(1-L_V(z))}{\EX{V}(1-z)}.
\end{equation}
Substituting $z = (1-\gamma)$ in the above expression, and differentiating, we get
\begin{equation}
\begin{split}
L_T(1-\gamma) & = \frac{(1-\rho)(1-L_V(1-\gamma))}{\gamma \EX{V}}, \text{ and }\\
\frac{d}{dz} L_T(z) \bigg|_{z = 1-\gamma} & = \frac{(1-\rho)}{\gamma \EX{V}}\bigg( \frac{1-L_V(1-\gamma)}{\gamma L_s(1-\gamma)} -L'_V(1-\gamma)\bigg).
\end{split}
\end{equation}
Using \eqref{pf:bgv:eq1}, we get
\begin{equation}
\begin{split}
A^\text{ave}  \geq &\frac{2(1-\rho)}{\gamma^2 \EX{V}} \Bigg(\bigg(2-\gamma + \frac{1}{L_S(1-\gamma)}\bigg)(1-L_V(1-\gamma)) \\
& - \gamma L'_V(1-\gamma)  \Bigg) + \frac{1}{2} - \frac{1}{\gamma}+ \frac{2}{\mu} \\
& + \frac{\gamma \EX{S^2} - \rho}{2(1-\rho)}  + \frac{\EX{V^2}}{2\EX{V}} \triangleq A^{\text{ave}}_{\text{LB}}.
\end{split}
\end{equation}
Similarly, using $\lceil a \rceil \leq a+1$ and simplifying as before, we get the corresponding upper bound
\begin{equation}
\begin{split}
A^\text{ave}&  \leq  A^{\text{ave}}_{\text{LB}} + (1-\rho)\frac{(1-L_V(1-\gamma))}{\gamma} \\
& + (1-\gamma)(1-\rho)\bigg(\frac{1-L_V(1-\gamma)}{\gamma L_S(1-\gamma)} - L'_V(1-\gamma) \bigg) \triangleq A^{\text{ave}}_{\text{UB}}.
\end{split}
\end{equation}

\end{document}